\documentclass[prb,twocolumn,showpacs,superscriptaddress,groupedaddress,amsmath,amssymb]{revtex4-1}

\usepackage{graphicx}
\usepackage{dcolumn}
\usepackage{bm}
\usepackage{hyperref}
\usepackage{booktabs}
\usepackage{float}

\begin{document}

\title{
Orbital-Selective Coexistence of Interlayer Spin-Singlet Formation and SDW Order with Anomalous Spin Reconfiguration in Bilayer Nickelate La$_{3}$Ni$_{2}$O$_{7}$ \\  Revealed by $^{17}$O-NMR
}

\author{Hongnam Lee$^1$, Mitsuharu Yashima$^1$, Masataka Kakoi$^{1,3}$, Takumi Ino$^1$, Yutaro Arai$^2$, Kentaro Kitagawa$^2$, \\Hiroya Sakurai$^{4}$, Yoshihiko Takano$^{4,5}$, Kazuhiko Kuroki$^{3}$, and Hidekazu Mukuda$^{1}$}

\email{email : mukuda.hidekazu.es@osaka-u.ac.jp}

\affiliation{$^1$Graduate School of Engineering Science, The University of Osaka, Toyonaka, Osaka 560-8531, Japan}
\affiliation{$^2$Institute for Solid State Physics (ISSP), The University of Tokyo, Kashiwa, Chiba 277-8581, Japan}
\affiliation{$^3$Department of Physics, The University of Osaka, Toyonaka, Osaka 560-0043, Japan}
\affiliation{$^4$National Institute for Materials Science (NIMS), Tsukuba, Ibaraki 305-0047, Japan}
\affiliation{$^5$Graduate School of Pure and Applied Sciences, University of Tsukuba, Tsukuba, Ibaraki 305-8577, Japan}

\date{\today}

\begin{abstract}
The spin structure of the spin density wave (SDW) order in the bilayer nickelate La$_3$Ni$_2$O$_7$ has been investigated using site-selective $^{17}$O-NMR measurements on the inner apical O(1), outer apical O(2), and planar O(3,4) sites.
Below $T_{\rm SDW}$ (= 150 K), the peak of all planar O(3,4) sites significantly broadens due to the emergence of a finite internal magnetic field, whereas O(2) sites remain with no (or a negligibly small) internal field. 
These results are consistent with commensurate SDW order with a single spin-spinless (or large-tiny spin) stripe. 
As for the O(1) sites that bridge the NiO$_2$ planes, the internal field is nearly canceled below $T_{\rm SDW}$, indicating an antiparallel spin configuration between adjacent planes.
However, below $T_\text{A}$ ($\sim$ 115 K), the spectrum of the O(1) site disappears even though the in-plane SDW order remains robust, implying that the antiparallel spin configuration through the Ni--O(1)--Ni bond is not particularly stable below $T_{\rm A}$, despite the expected strong interlayer spin coupling between the NiO$_2$ planes. 
Above all, we emphasize that the local spin susceptibility is extremely small at the O(2) site that has a strong covalency with the $d_{3z^2-r^2}$ orbital, indicating a well-developed interlayer spin-singlet formation in the Ni-$d_{3z^2-r^2}$ orbitals bridging the NiO$_2$ planes.
These findings shed new light on the interlayer spin-singlet formation and the anomalous spin reconfiguration through the $\text{Ni--O(1)--Ni}$ bonding orbitals connecting the NiO$_2$ planes, which characterize the orbital-selective nature of the bilayer nickelate La$_3$Ni$_2$O$_7$.
\end{abstract}

\maketitle

\setcounter{section}{0}
\renewcommand{\thesection}{\arabic{section}} 
\section{\leftline{I\lowercase{ntroduction}}}

In 2023, high-temperature superconductivity with the superconducting (SC) transition temperature $T_{\rm c}\sim 80$~K was discovered in the bilayer nickel oxide La$_3$Ni$_2$O$_{7}$ under high pressure\cite{Sun_2023}.
The spin and charge degrees of freedom of the electrons in $d_{x^{2}-y^{2}}$ and $d_{3z^{2}-r^{2}}$ orbitals are expected to play a crucial role in the occurrence of the SC state at high pressures.
At ambient pressure, anomalies in the electronic states were reported below 150 K by many experimental studies on La$_3$Ni$_2$O$_{7+\delta}$ (La327), which have been attributed to the emergence of charge density wave (CDW) and/or spin density wave (SDW) orders\cite{Zhang, Wu, Liu2023, Liu2024, Hou_replication_2023, Yanan-Zhang_replication, G-Wang_2024, Chen_RIXS, Taniguchi, Kakoi, Meng, Chen_muSR, Khasanov, Ren, Gupta, Zhao_NMR,Yashima327,Plokhikh,Luo}. 
Among the possible candidates, the single spin-charge stripe order\cite{Chen_RIXS,Chen_muSR,Khasanov,Ren,Yashima327,Plokhikh}, the double spin stripe order\cite{Chen_RIXS,Gupta,Zhao_NMR}, the double charge stripe order\cite{Luo}, and the double spin-charge stripe order have been discussed in various spectroscopic experiments. 
Nevertheless, a consensus has yet to be reached. 
To further resolve the spin and charge states in the ordered phase, it is important to consider that this compound tends to contain some defects such as oxygen deficiencies and stacking faults. 
Therefore, it is necessary to carefully determine whether the observed physical properties are intrinsic or extrinsic. 
This motivated us to perform a $^{139}$La-nuclear quadrupole resonance (NQR) study on La327, which suggested commensurate SDW order with a single-charge (spinless) stripe by selectively probing the La(2)$_a$ sites of an ideal La327 structure\cite{Yashima327}.
As the La(2) site probes the NiO$_2$ plane from outside the bilayer blocks, it was not possible to directly identify the entire spin structure between the NiO$_2$ planes.
Therefore, investigating the local electronic states at the oxygen sites using $^{17}$O-NMR is more effective to unveil the microscopic spin and charge states of the bilayered NiO$_2$ planes.

In this paper, we report on a site-selective $^{17}$O-NMR measurement in the SDW phase of the bilayer nickelate La$_3$Ni$_2$O$_7$ at ambient pressure. 
Three oxygen sites are distinguished by utilizing the differences in their local microscopic states. 
The results on these O sites are consistent with commensurate SDW order with a single spin-spinless (or large-tiny spin) stripe below $T_{\rm SDW}$ = 150 K. 
The interlayer spin configuration between the NiO$_2$ planes is antiparallel just below 150 K; however, it turns non-antiparallel below $T_\text{A} \sim$ 115 K, indicating that the interlayer antiparallel spin configuration is not as robust as anticipated, despite the expected strong interlayer spin coupling between the NiO$_2$ planes. 
We note that the local spin susceptibility ($\chi_{\rm spin}^{\rm loc}$) at the O(2) site is extremely small, indicating a well-developed interlayer spin-singlet formation between the Ni-$d_{3z^2-r^2}$ orbitals that bridge the NiO$_2$ planes.
We discuss the anomalous interlayer coupling effects between NiO$_2$ planes driving the interlayer spin-singlet formation, as well as the spin reconfiguration within the SDW phase, both of which highlight the orbital-selective nature of the bilayer nickelate.

\section{\leftline{E\lowercase{xperimental}}}

Polycrystalline La$_3$Ni$_2$O$_{7+\delta}$ ($\delta\simeq0.0$) was prepared using the solid-state reaction method described elsewhere.\cite{Ueki}
The oxygen content was estimated by thermogravimetric analysis. 
The exchange of natural oxygen $^{16}$O ($I$ = 0) for $^{17}$O ($I$ = 5/2) was carried out by annealing the sample at 1000~$^\circ$C for 4 hours.
The high quality of the sample after the $^{17}$O substitution was confirmed by the observation of the very narrow $^{139}$La(2)-NQR spectrum.
No noticeable differences were observed compared to the high-quality sample ($\delta\simeq0.0$) reported previously, which has been microscopically guaranteed by NQR probes to date\cite{Yashima327}.
A coarse powder sample was used for the $^{17}$O-NMR experiment, performed by sweeping the frequency of radio-frequency pulses ($f_{\rm RF}$) under an external magnetic field of $H_0$ = 11.968 T. 
The NQR frequency is defined as $\nu_Q^{(i)} = 3eQV_{zz}^{(i)} \!/ [2I(2I-1)h]$, where $V_{zz}^{(i)}$ is a principal value of the electric field gradient (EFG) tensor at each O($i$) site, and $Q$ is the electric quadrupole moment of $^{17}$O. 
The asymmetry parameter of the EFG $\eta$ is defined as $|V_{xx}-V_{yy}|/V_{zz}$. 
The calculation of $\nu_Q^{\rm cal}$ and $\eta^{\rm cal}$ for each O site was performed with the \textsc{WIEN2k} code\cite{Wien2K} within density functional theory using the generalized gradient approximation with the Perdew-Burke-Ernzerhof parametrization\cite{Perdew} and the full-potential linearized augmented plane-wave method, based on the experimental lattice parameters\cite{G-Wang_2024}. 
The results are summarized in Table I.

\begin{table}[H]
\centering
\caption{
Site assignment of three $^{17}$O-NMR peaks.
The experimental intensity ratio (IR$^{\rm exp}$) for each O site is evaluated relative to the intensity for the O(1) site based on the analysis of the observed spectrum. These ratios are close to the site number ratio (NR) relative to that of the O(1) site. 
The NQR frequency ($\nu_Q^{\rm exp}$) at each site determined experimentally is also similar to the values obtained by calculation ($\nu_Q^{\rm cal}$). 
}
\label{tab:nqr_parameters}

\setlength{\heavyrulewidth}{0.85pt}
\setlength{\lightrulewidth}{0.60pt}
\renewcommand{\arraystretch}{1.15}
\resizebox{\linewidth}{!}{%
\begin{tabular}{lcccccc}
\toprule
Peak
& Site
& NR
& IR$^{\rm exp}$
& $\nu_Q^{\rm exp}$
& $\nu_Q^{\rm cal}$
& $\eta^{\rm cal}$ \\
&
&
&
&
(MHz)
& (MHz)
& \\
\midrule
1 & O(1)   & 1 & 1
& $\sim$0.68 & 1.17 & 0.032 \\
2 & O(2)   & 2 & $\sim$1.6
& $\sim$0.10 & 0.17 & 0.23 \\
3 & O(3,4) & 4 & $\sim$4.0
& $\sim$0.78 & 0.907, 0.912 & 0.16, 0.13 \\
\bottomrule
\end{tabular}
}
\end{table}
\section{Results and Discussion}

\subsection{Assignment of three $^{17}$O sites above $T_{\rm SDW}$ }

\begin{figure}[htbp]
\hspace*{-0cm}
\includegraphics[width=7.5cm]{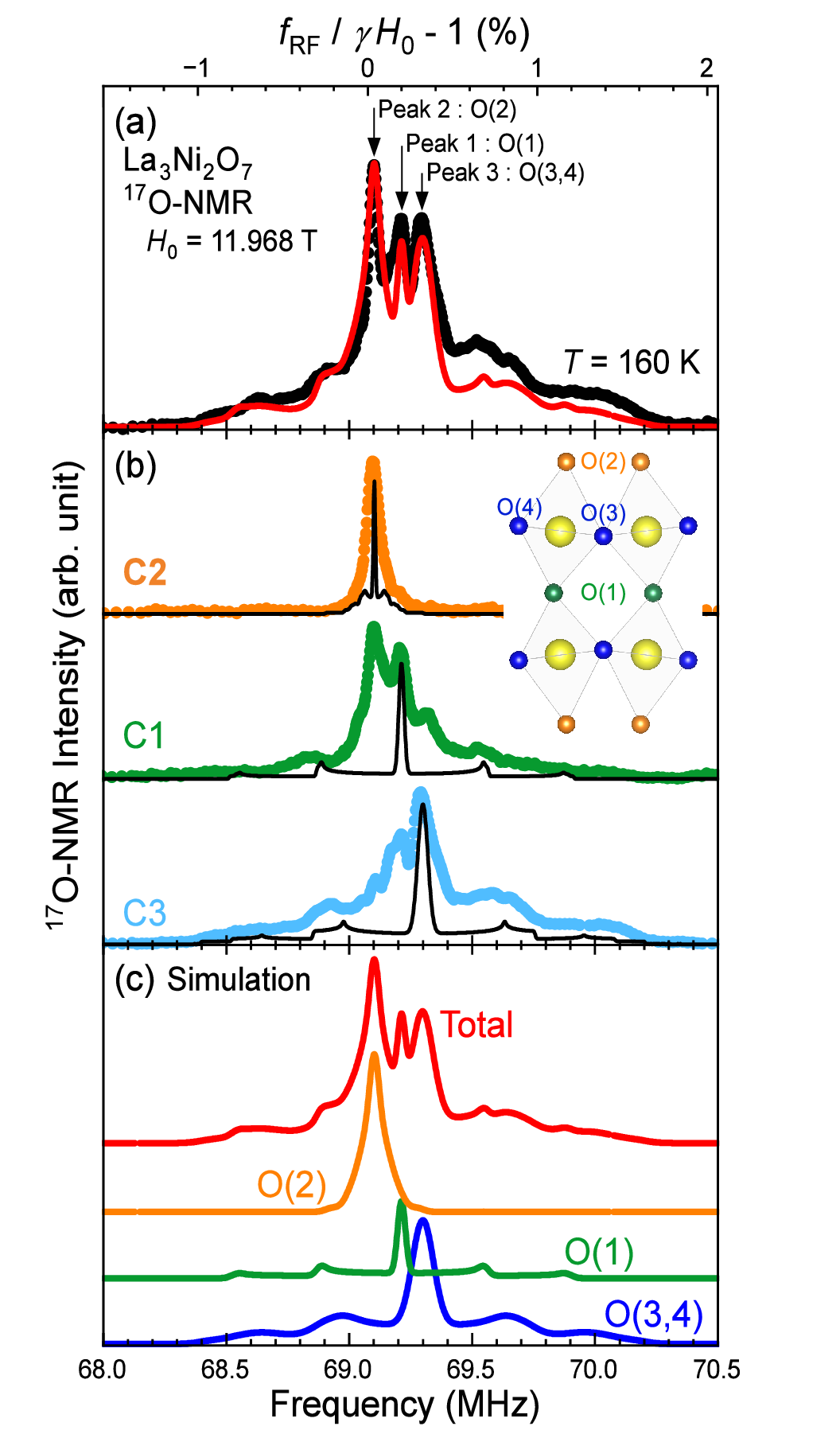}
\vspace*{-0.2cm}
\caption[]{\footnotesize (color online) (a) $^{17}$O-NMR spectrum at 160 K ($>T_{\rm SDW}$). The red curve is the simulated result shown in (c). 
Here, the broad background spectral component from the $^{139}$La-NMR is subtracted.
The top axis indicates the shift relative to $\gamma H_0$ ($\gamma$ is the gyromagnetic ratio of $^{17}$O), where the zero point corresponds to the position of $K = 0$ for $^{17}$O nuclear spins.
(b) Three peaks are distinguished using three different NMR pulse conditions (C1, C2, and C3), enabling the assignment of the outer apical O(2), inner apical O(1), and planar O(3,4) sites, respectively. 
The black curves are simulated spectra to determine the experimental values of $\nu_Q^{\rm exp}$ for the three O sites. 
(c) The simulation of the entire NMR spectrum (red curve) is obtained by superposing the components simulated for the three O sites using their $\nu_Q^{\rm exp}$ and IR$^{\rm exp}$ (see Table I) with an appropriate spectral broadening. 
All these results demonstrate the unambiguous site-assignment of the three O sites.
}
\label{Spec160K}
\end{figure}

Figure \ref{Spec160K}(a) shows the $^{17}$O-NMR spectra at 160 K above $T_{\rm SDW}$ (= 150 K). 
Three distinct peaks are observed, which are labeled peak~2, peak~1, and peak 3 from low to high frequency. 
Ideally, there are four oxygen sites in orthorhombic La$_3$Ni$_2$O$_{7}$: inner apical O(1), outer apical O(2), and planar O(3,4) sites, as shown in the inset of Fig.~\ref{Spec160K}(b). 
To identify the origin of the three peaks, we utilized the differences in the longitudinal and transversal nuclear spin relaxation times ($T_1$ and $T_2$) for each O site, since both are the longest for peak~2 and the shortest for peak~3.
In general, the intensity of a site with short $T_2$ can be suppressed by a long time interval ($\tau$) between the $\pi/2$ and the $\pi$ pulses, and the intensity of a site with long $T_1$ can be suppressed by a short repetition time (TR), which is the time interval between successive $\pi/2-\pi$ pulse sequences, in the spin-echo method. 
To selectively observe peak~2, we measured the spectrum under the NMR pulse condition C2 ($\tau$ = 1~ms, TR = 290~ms), as shown in Fig.~\ref{Spec160K}(b). 
The observation of the narrow spectrum indicates that peak 2 derives from the site with the smallest NQR frequency of $\nu_Q^{\rm exp}(2) \sim 0.1$~MHz, which is determined by comparison with the simulation (black curve). 
Using pulse condition C3 ($\tau = 12$ $\mu$s, TR = 20 ms), the central and satellite peaks of peak 3 can be clearly distinguished, as shown in Fig.~\ref{Spec160K}(b), indicating that this site has a relatively large NQR frequency of $\nu_Q^{\rm exp}(3)\sim 0.78$~MHz. 
Using the intermediate pulse condition C1 ($\tau = 400~\mu$s, TR = 20~ms), the spectral intensity of peak~1 is relatively dominant compared with the other peaks, as shown in Fig.~\ref{Spec160K}(b), indicating that peak~1 originates from the site with $\nu_Q^{\rm exp}(1) \sim 0.68$~MHz. 
Here, the broad background spectral component from the $^{139}$La-NMR in Fig.~\ref{Spec160K}(b) can be almost eliminated by using a moderately adjusted $\pi/2$ pulse added to the NMR pulse sequence mentioned above.
To assign the O sites of these peaks, we compare the experimental values of $\nu_Q^{\rm exp}$ with the $\nu_Q^{\rm cal}$ estimated by \textsc{WIEN2k} for each O site, as summarized in Table \ref{tab:nqr_parameters}.
Peak 2 is assigned to the outer apical O(2) site, since the $\nu_Q^{\rm exp}$ and $\nu_Q^{\rm cal}$ values are the smallest and particularly similar. 
Peaks 1 and 3 could not be distinguished solely by their $\nu_Q$ values, since the $\nu_Q^{\rm cal}$ and $\nu_Q^{\rm exp}$ values resemble each other. 
However, the experimental NMR intensity ratio (IR$^{\rm exp}$) of peak 1 to peak 3 is approximately 1 : 4.0, which coincides with the ideal site number ratio (NR) of 1 : 4 for the O(1) to O(3,4) sites, where IR$^{\rm exp}$ for each O site is evaluated relative to the intensity for the O(1) site based on the analysis of the observed spectrum. 
This excellent agreement allows the assignment of peak~1 and peak~3 to the inner apical O(1) and the planar O(3,4) sites, respectively.
The spectral components for each O site are simulated using the $\nu_Q^{\rm exp}$ and $\eta^{\rm cal}$ values, as shown in Fig.~\ref{Spec160K}(c).
Consequently, the entire spectrum can be well reproduced, as shown by the red curves in Figs.~\ref{Spec160K}(a) and \ref{Spec160K}(c), by superposing the simulations of three O sites with the intensity ratio (IR(1)$^{\rm exp}$ : IR(2)$^{\rm exp}$ : IR(3,4)$^{\rm exp}$ $\approx 1 : 1.6 : 4.0$), which is very close to the ideal NR. 
Here, we note that in this NMR measurement of the polycrystalline sample, the planar O(3) and O(4) sites cannot be distinguished experimentally due to the similarity of their local NQR parameters.

\subsection{Insights from the $^{17}$O-Knight shift}

Next, we focus on the Knight shift ($K$) at each O site: The $K$ values at 160~K are  $\sim$0.04 \% for O(2), $\sim$0.20 \% for O(1), and $\sim$0.33 \% for O(3,4), as shown in Fig. \ref{Spec160K}(a). 
In general, the observed $K$ is composed of an orbital part ($K_{\rm orb}$) and a spin part  ($K_{\rm spin}$) that is proportional to the $\chi_{\rm spin}^{\rm loc}$ at each O site.
Although $K_{\rm orb}$ cannot be precisely identified at present, from the previous $^{17}$O-NMR studies on the related oxides, it is deduced to be negligibly small compared to the observed $K$: For example, $K_{\rm orb}$ at the O(2) site is nearly zero in the monolayer nickelate La$_2$NiO$_4$\cite{Halat}, and less than $\lvert 0.004 \rvert~\%$ in the bilayer ruthenate Sr$_3$Ru$_2$O$_7$\cite{Kitagawa327}. 
Thus, we reasonably infer that the observed $K$ in La327 approximately corresponds to $K_{\rm spin}$, which is dominated by the $\chi_{\rm spin}^{\rm loc}$ at the O-$2p$ orbital polarized through the hybridization with the neighboring Ni sites. 
The $K$ value at the O(2) site is one order of magnitude smaller than that at the planar O(3,4) sites, indicating the large reduction of the $\chi_{\rm spin}^{\rm loc}$ at the O(2) site in spite of the substantial hybridization between O(2)-$p_z$ and Ni-$d_{3z^2-r^2}$ orbitals. 
The small $\chi_{\rm spin}^{\rm loc}$ at the O(2) site is also corroborated by the fact that the $T_1$ and $T_2$ values at the O(2) site are much longer than at the O(3,4) site, as discussed in the C2 condition of Fig.~\ref{Spec160K}(b). 
The large reduction of the $\chi_{\rm spin}^{\rm loc}$ at the O(2) site in La327 strongly suggests the well-developed interlayer spin-singlet formation between the Ni-$d_{3z^2-r^2}$ orbitals bridging the NiO$_2$ planes through the O(1) site. 
This is in contrast to the case of monolayer La$_2$NiO$_4$ with no interlayer coupling, in which the larger $\chi_{\rm spin}^{\rm loc}$ at the O(2) site is manifested by a substantial $K_{\rm spin}$ of $\sim0.35\%$~\cite{Halat}. 
In La327, the spin of the Ni-$d_{3z^2-r^2}$ orbitals is nearly quenched by the prominent spin-singlet formation, resulting in an extremely small $\chi_{\rm spin}^{\rm loc}$ at the O(2) site located along the $c$-axis. 
Instead, the Ni spin resides primarily in the $d_{x^2-y^2}$ orbital, which hybridizes extensively with the neighboring O(3,4)-$p_{x/y}$ orbitals within the $ab$-plane, thereby leading to a large $\chi_{\rm spin}^{\rm loc}$ ($\propto K_{\rm spin}$) at the O(3,4) sites. 
These findings clearly indicate the orbital-selective nature of the Ni spin state in the bilayer nickelate La327, where the $d_{3z^2-r^2}$ orbitals dominate the interlayer spin-singlet formation while the planar $d_{x^2-y^2}$ orbitals carry the vast majority of the Ni spin. Such a strong spin-singlet formation likely originates from the interlayer bonding between the NiO$_2$ planes, which is strong enough to induce a pronounced bonding--antibonding splitting of the $d_{3z^2-r^2}$ orbital states.

Here, we note that the $K$ at the O(1) site in this bonding is moderately larger than that at the O(2) site, which can be attributed to two main factors. 
First, the O(1) site is sandwiched between two Ni sites along the $c$-axis, whereas the O(2) site is coordinated to only one Ni site. 
Second, the shorter Ni--O(1) distance than the Ni--O(2) distance enhances the transferred hyperfine field at the O(1) site. 
Regarding the possible origins of the transferred hyperfine fields at the O(1)/O(2) sites, we mainly consider three distinct types of hybridization of the O(1)/O(2)-$p_z$ orbital with: (i) Ni-$d_{3z^2-r^2}$, (ii) Ni-$d_{x^2-y^2}$, and (iii) O(3,4)-${p_x}/{p_y}$. 
In path (i), the spin of the Ni-$d_{3z^2-r^2}$ orbital is predominantly extinguished by the substantial interlayer spin-singlet formation, leaving almost no transferred hyperfine field at the O(1)/O(2) sites. However, any slight residual spin component of this orbital remaining even under the well-developed spin-singlet formation could provide a tiny transferred hyperfine field. 
For path (ii), the bending of the Ni--O(1)--Ni bond angle is expected to induce a weak hybridization between the O-$p_z$ and Ni-$d_{x^2-y^2}$ orbitals, although no hybridization is expected in an ideal case of the straight coordination. 
It leads to the occurrence of a possible transferred hyperfine field at the O(1)/O(2) sites from the spins of the Ni-$d_{x^2-y^2}$ orbital. 
Path (iii) represents a transferred hyperfine field originating not from the Ni sites located directly above or below the O(1)/O(2) sites, but rather from neighboring Ni sites via the O(3,4)-$p_x$/$p_y$ orbitals.
Generally, a decrease in the distance between the Ni and O(1)/O(2) sites enhances the hybridization along all of these paths, leading to an increase in the transferred hyperfine fields. 
The converse is also true for an increase in the Ni--O(1)/O(2) distance.
Consequently, the larger $K$ at the O(1) site is attributed to the shorter Ni--O(1) distance that enhances the transferred field moderately at the O(1) site. 
In contrast, such paths become practically negligible at the O(2) site due to the longer distance, namely, the substantially weakened hybridization, resulting in a significantly smaller $K$.

\subsection{Spin structure of SDW state for $115 < T < 150$~K}              

\begin{figure}[t]
\hspace*{-0.0cm}
\includegraphics[width=8.5cm]{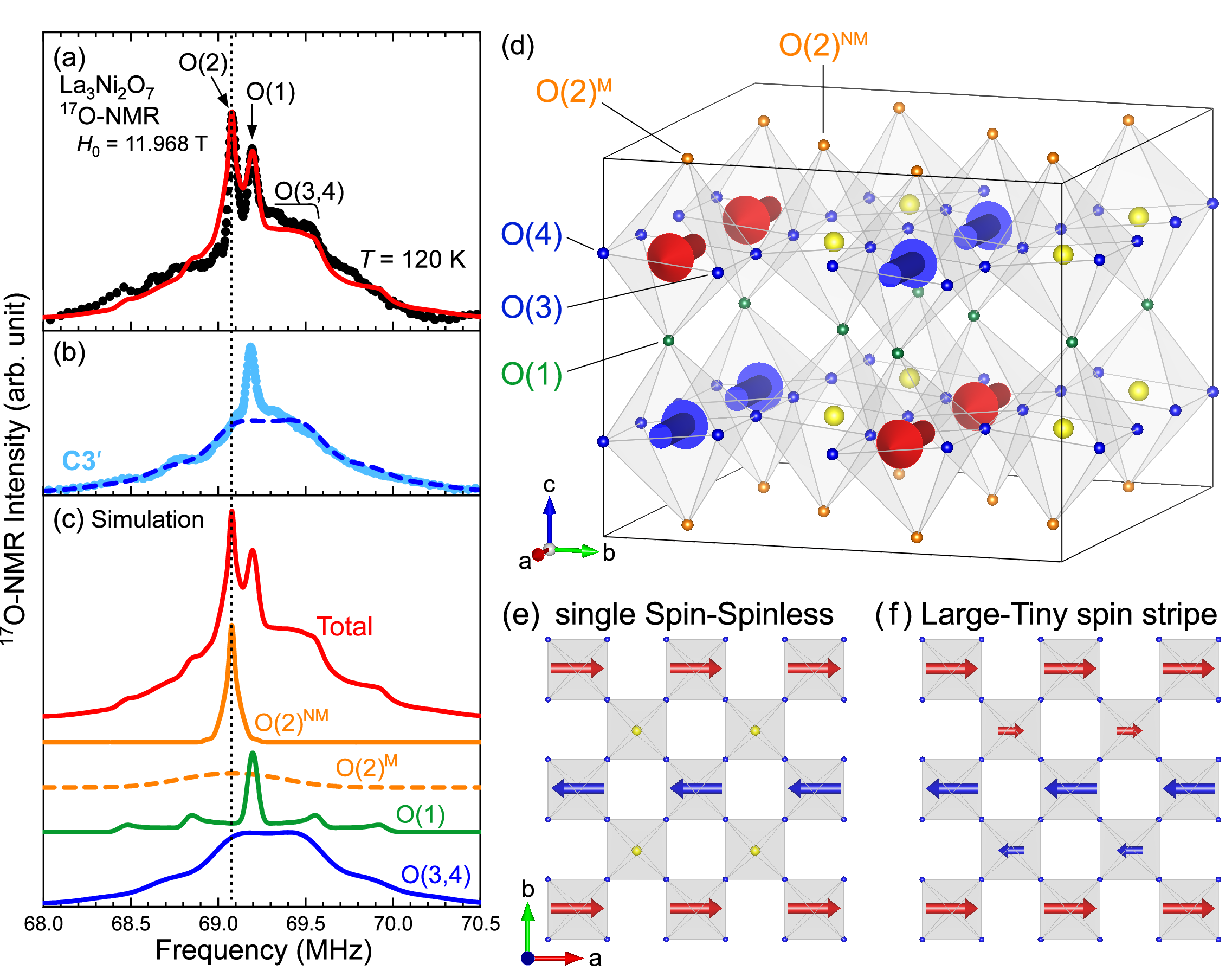}
\vspace*{-0.5cm}
\caption[]{\footnotesize (color online) (a) $^{17}$O-NMR spectrum of the SDW state at 120 K ($T_{A}\le T \le T_{\rm SDW}$). 
The peak of the O(3,4) sites is largely broadened due to a finite $H_{\rm int}$ from the adjacent Ni moments in the commensurate SDW order. 
The broad spectral component from the $^{139}$La-NMR on the background is subtracted.
The vertical dotted line indicates the position of $K = 0$ for $^{17}$O nuclear spins.
(b) The NMR pulse condition (C3') can selectively highlight the spectrum of the O(3,4) sites, revealing the presence of $H_{\rm int}(3,4)\sim 0.05$ T from the simulation (broken curve).
(c) Spectrum simulations for the O(3,4) with the finite $H_{\rm int}(3,4)$, the O(2)$^{\rm NM}$ with no or a negligibly small $H_{\rm int}$ and the O(2)$^{\rm M}$ with finite $H_{\rm int}(2)^{\rm M}$, and the O(1) with no $H_{\rm int}$. 
The broken curve for the O(2)$^{\rm M}$ is a hypothetical component to reproduce the entire observed spectrum.
(d) The SDW model with the antiparallel interlayer spin configuration. It fairly reproduces the entire spectrum, as shown by the red curves in (a) and (c), which are obtained by superposing those spectral components with an intensity ratio of IR(1) : IR(2)$^{\rm NM}$ : IR(2)$^{\rm  M}$ : IR(3,4) $\approx 1:0.5:0.5:4$, which is reasonably close to the ideal one ($1:1:1:4$).
Spin structures of the in-plane SDW phase with (e) the single spin-spinless stripe and (f) the large-tiny spin stripe.
}
\label{Spec120K}
\end{figure}

\begin{figure*}[htbp]
\hspace*{-0cm}
\includegraphics[width=17cm]{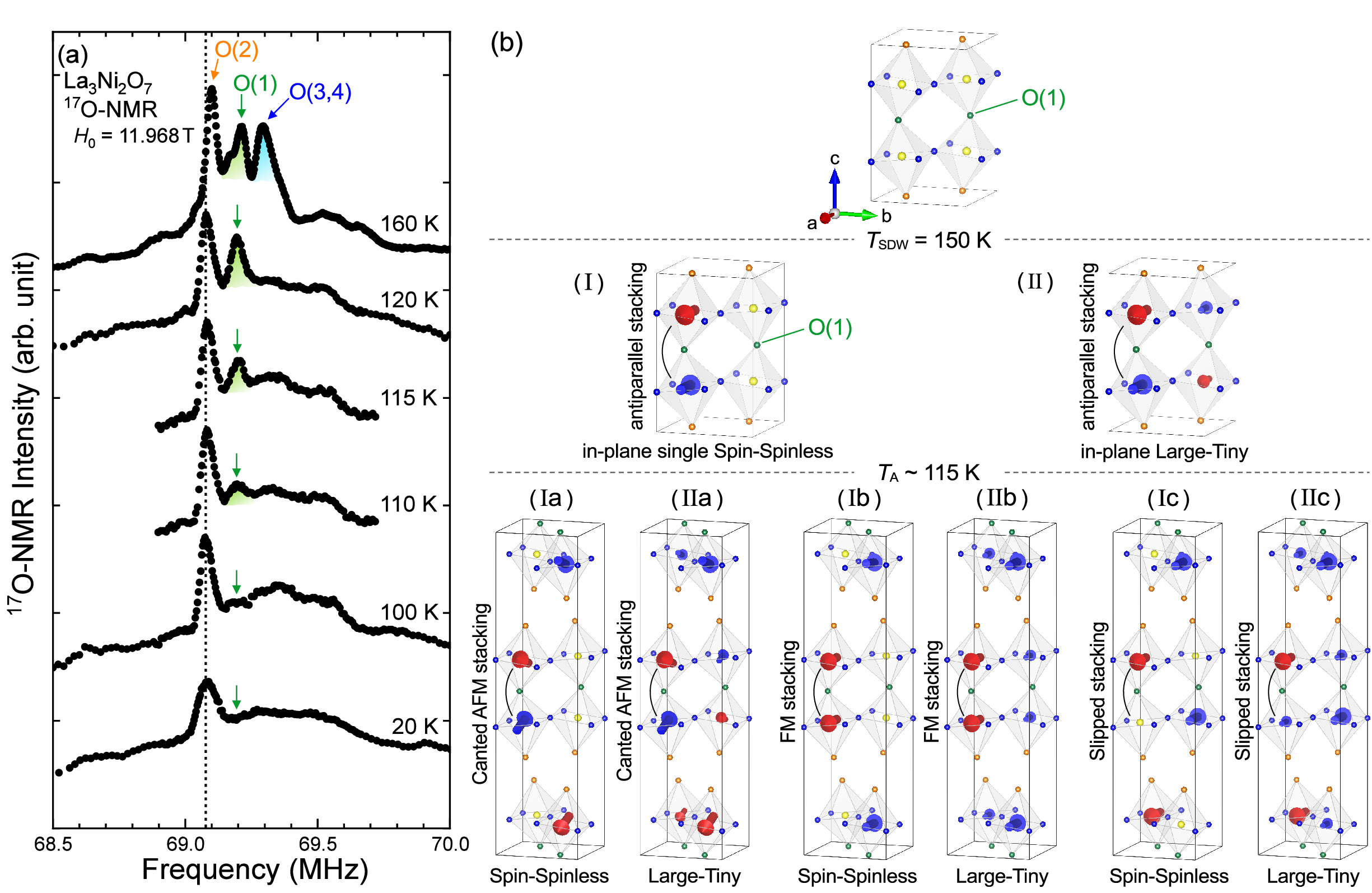}
\vspace*{-0.2cm}
\caption[]{\footnotesize (color online) (a) $T$ dependence of $^{17}$O-NMR spectra. 
Among the three narrow peaks of each O site, the peak associated with the planar O(3,4) sites broadens significantly below $T_{\rm SDW}$ (= 150 K). 
As for the inner apical O(1) sites, the narrow peak is present from $T_{\rm SDW}$ to 115 K ($\equiv T_{\rm A}$), suggesting that the interlayer spin configuration is predominantly antiparallel.
However, this O(1) peak disappears drastically below $T_{\rm A}$, even though the in-plane SDW order does not change, implying that the interlayer spin configuration between adjacent SDW planes becomes non-antiparallel.
The vertical dotted line indicates the position of $K = 0$ for $^{17}$O nuclear spins.
(b) The possible stacking configurations along the $c$-axis are schematically illustrated for two in-plane SDW types, the spin-spinless stripe (I) and large-tiny spin stripe (I\hspace{-1.2pt}I).
The possible candidates are the canted AFM stacking (Ia, I\hspace{-1.2pt}Ia), ferromagnetic stacking (Ib, I\hspace{-1.2pt}Ib), and slipped stacking (Ic, I\hspace{-1.2pt}Ic). 
We suggest that the canted AFM stacking is the most viable candidate among them (see text).
}
\label{Spec20K}
\end{figure*}

Next, we discuss the $^{17}$O-NMR spectra under the SDW state at $T$ = 120 K in Fig. \ref{Spec120K}(a). 
In comparison with the spectra at 160 K (see Fig. \ref{Spec20K}(a)), the peak of the planar O(3,4) sites is completely broadened at 120 K ($<T_{\rm SDW}$), indicating that all of the O(3,4) sites experience a finite internal magnetic field ($H_{\rm int}$) from neighboring Ni spins. 
The presence of a narrow peak from the O(2) sites indicates that some O(2) sites experience no or a negligibly small $H_{\rm int}$ even in the SDW state.
These facts are consistently accounted for by the commensurate SDW order with a single spin-spinless stripe along the $b$-axis, as shown in Fig.~\ref{Spec120K}(e). 
We note that the present result can be also explained by the large-tiny spin stripe model\cite{Plokhikh} in Fig.~\ref{Spec120K}(f), if we assume the sufficiently tiny spin that induces a negligibly small $H_{\rm int}$ at the O(2) site. 
These correspond to the striped SDW structure with a magnetic moment along the $a$-axis\cite{Yashima327}. 
Among the other candidates of the SDW models, the double spin stripe ($\cdots \uparrow \,\, \uparrow\,\, \downarrow \,\, \downarrow \,\, \uparrow \,\, \uparrow \,\, \downarrow\,\, \downarrow \cdots$) is excluded, since half of the O(3,4) sites are expected to be nonmagnetic due to the cancellation of $H_{\rm int}$ at the O(3,4) sites between the Ni($\uparrow$) and Ni($\downarrow$) sites. 
Additionally, this model requires that all the O(2) sites be magnetic, which contradicts our observations. 
Similarly, the double spin-charge stripe ($\cdots \circ \circ \uparrow \, \uparrow \circ\, \circ \downarrow \, \downarrow \circ \circ \cdots$) is excluded, since one-fourth of the O(3,4) sites between the spinless Ni sites are expected to be nonmagnetic. 
Consequently, among the candidates of the SDW models, the single spin-spinless (or large-tiny spin) stripe model consistently explains the present $^{17}$O-NMR results, as well as the other previous spectroscopic experiments\cite{Chen_RIXS,Chen_muSR,Khasanov,Ren,Yashima327,Plokhikh}.
Here, the charge anomaly was not explicitly detected across $T_{\rm SDW}$ in the measured $T$ region, but a slight CDW order below the detection limit of the NQR/NMR methods cannot be ruled out.

Next we analyze the spectral components of three O sites. 
The spectrum in Fig. \ref{Spec120K}(b) was obtained using the NMR pulse condition (C3') with a shorter TR (= 50 ms) than the TR (= 1.5 s) used in Fig. \ref{Spec120K}(a), which can selectively highlight the broad spectrum of the O(3,4) sites by eliminating the intensity of the O(2) site in particular.
This spectrum of O(3,4) sites can be reproduced by assuming $H_{\rm int}(3,4) \sim 0.05$ T, as shown by the simulation (the broken curve) in Fig. \ref{Spec120K}(b).
Regarding the outer apical O(2) sites, nearly half of them exhibit a negligibly small $H_{\rm int}$ (nonmagnetic O(2)$^{\rm NM}$), whereas the other half on the spin channels experience a sizable $H_{\rm int}(2)^{\rm M}$ (magnetic O(2)$^{\rm M}$) from an adjacent Ni moment, as illustrated in Fig.~\ref{Spec120K}(d).
The $H_{\rm int}(2)^{\rm M}$ at the O(2)$^{\rm M}$ sites could not be experimentally determined, and thus here we tentatively assumed that it is similar to the value at O(3,4) sites.
As for the O(1) sites that bridge the NiO$_2$ planes, the spectrum mostly remains narrow, implying that the $H_{\rm int}$ at O(1) sites is almost canceled. 
Assuming the SDW model with an antiparallel spin configuration between the adjacent planes along the $c$-axis, as shown in Fig.~\ref{Spec120K}(d), even the O(1) sites between the Ni($\uparrow$) and Ni($\downarrow$) sites are also non-magnetic due to the cancellation of $H_{\rm int}$ from antiparallel Ni spins on the adjacent planes.
As a result, the simulations of the spectral components of the O(1), O(2)$^{\rm NM}$, O(2)$^{\rm M}$, and O(3,4) sites are summarized in Fig.~\ref{Spec120K}(c). 
The entire spectrum can be roughly reproduced, as shown by the red curves in Figs.~\ref{Spec120K}(a) and \ref{Spec120K}(c), by superposing those spectral components 
with the intensity ratio of IR(1) : IR(2)$^{\rm NM}$ : IR(2)$^{\rm M}$ : IR(3,4) $\approx$ $1 : 0.5 : 0.5 : 4$, which is reasonably close to the ideal one ($1 : 1 : 1 : 4$) expected from the SDW model with the antiparallel interlayer spin stacking along the $c$-axis. 

\subsection{Anomalous interlayer spin configuration below $T_{A}$ ($\sim$ 115 K)}              

Figure \ref{Spec20K} shows the $T$ dependence of the $^{17}$O-NMR spectra down to 20 K. 
We observed the pronounced reduction of the O(1) peak most likely caused by the significant spectral broadening below 115~K ($\equiv T_{A}$). 
This indicates that the $H_{\rm int}(1)$ is not fully canceled at all the O(1) sites below $T_{A}$. 
We emphasize that the commensurate SDW order within each NiO$_2$ plane does not change drastically, since no anomaly was detected across $T_{A}$ in the current O(2)-NMR and the previous La(2)-NQR measurements\cite{Yashima327}, both of which observe the in-plane SDW ordered state from outside the bilayer blocks. 
Therefore, this anomaly at $T_{\rm A}$ is attributed to the change in the spin configuration between the NiO$_2$ planes being no longer antiparallel.
As shown in Fig. \ref{Spec20K}(b), the possible interlayer stacking patterns along the $c$-axis are considered for the two types of in-plane striped SDWs: (I) spin-spinless stripe and (I\hspace{-1.2pt}I) large-tiny spin stripe. 
For each type, we discuss the three possible candidates: the canted antiferromagnetic (AFM) stacking (Ia and I\hspace{-1.2pt}Ia), the ferromagnetic (FM) stacking (Ib and I\hspace{-1.2pt}Ib), and the in-plane slipped (IS) stacking (Ic and I\hspace{-1.2pt}Ic).
Here, the IS stacking corresponds to a configuration in which the in-plane SDW spin pattern is shifted by ($\pm$1/2, $\pm$1/2) in real space relative to the adjacent NiO$_2$ plane.

\begin{figure}[htbp]
\hspace*{-0cm}
\includegraphics[width=7.0cm]{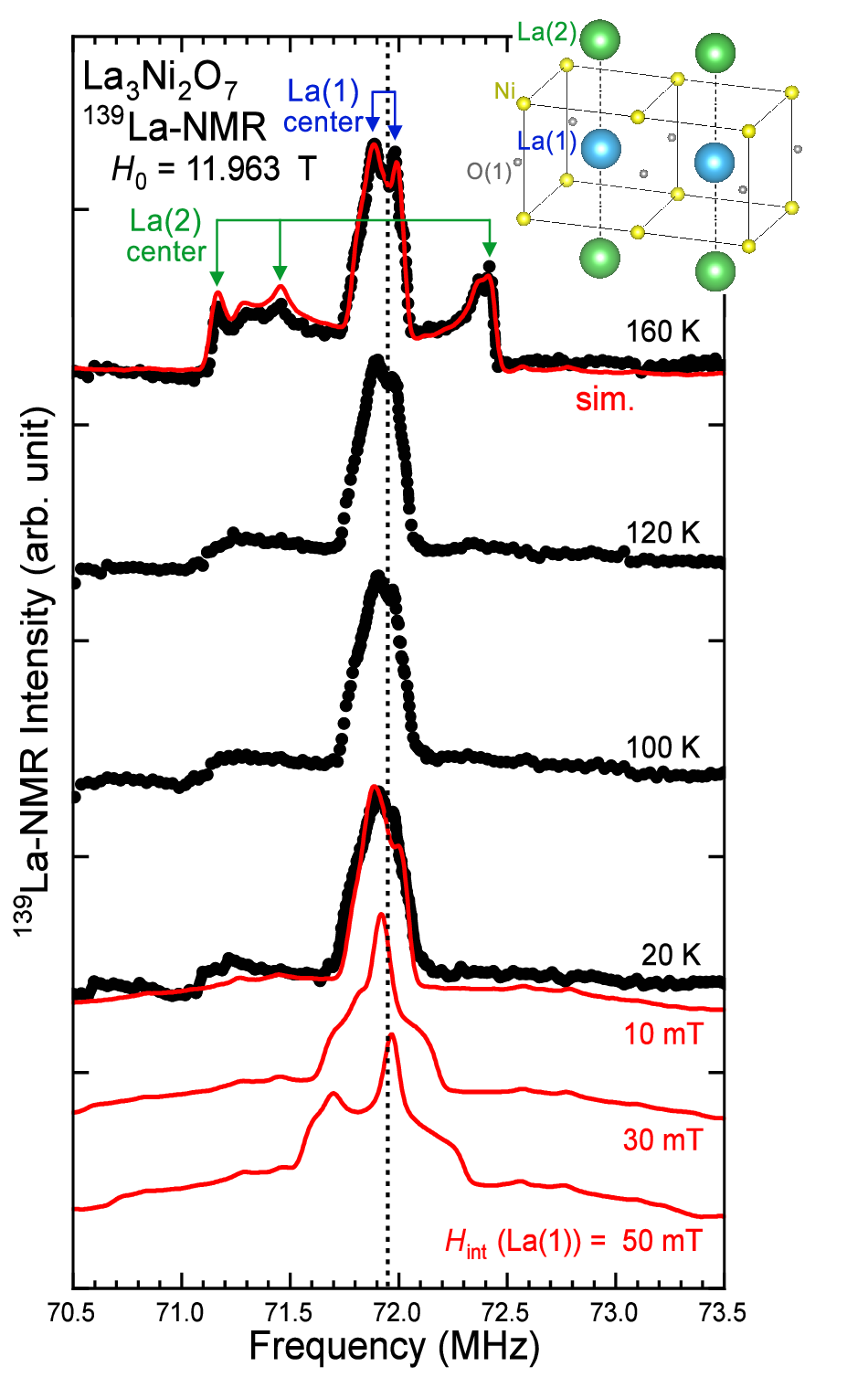}
\vspace*{-0.3cm}
\caption[]{\footnotesize
(color online) $T$ dependence of the $^{139}$La-NMR spectra of $^{17}$O-substituted La327 used in this experiment, which exhibit a superposition of randomly oriented powder patterns from the La(1) and La(2) sites.
At 160 K ($>T_{\rm SDW}$), the combined spectrum for the La(1) and La(2) sites is perfectly reproduced by the simulation of a randomly oriented powder pattern based on the previously reported parameters (the top red curve)\cite{Kakoi,Yashima327}.
Below $T_{\rm SDW}$, the La(2) spectral component shows a significant broadening due to the emergence of a large $H_{\rm int}$\cite{Yashima327}, whereas the La(1) component exhibits only a marginal broadening, which can be simulated by a sufficiently small $H_{\rm int}$ of at most $\sim$ 10~mT even well below $T_{\rm SDW}$. 
To evaluate this minimal broadening, the lower part of the figure displays the simulations assuming $H_{\rm int} \sim$ 10, 30, and 50~mT (red curves), compared with the experimental spectrum at 20 K.
Comparisons of the observed spectra with the anticipated hyperfine fields at the La(1) site for each stacking pattern suggest the canted AFM stacking rather than the FM or IS stacking as a compelling candidate for the spin reconfiguration below $T_{\rm A}$ (see text).
The vertical dotted line indicates the position of $K = 0$ for $^{139}$La nuclear spins.
}
\label{La1NMR}
\end{figure}

To identify the actual stacking configuration, the $H_{\rm int}$ at La(1) site between NiO$_2$ planes provides a complementary clue. 
Figure \ref{La1NMR} shows the $T$ dependence of $^{139}$La-NMR spectra for the polycrystalline sample used in this experiment, which exhibit a superposition of randomly oriented powder patterns from the La(1) and La(2) sites.
The central lines ($+1/2\Leftrightarrow-1/2$) for the La(1) and La(2) sites are clearly distinguished at 160~K ($>T_{\rm SDW}$).
Although the spectral component of the La(2) site shows the significant broadening due to the large $H_{\rm int}$ below $T_{\rm SDW}$\cite{Yashima327}, the La(1)-spectral component exhibits only a marginal broadening below $T_{\rm SDW}$. 
Assuming $H_{\rm int} \perp V_{zz}$ for the La(1) site, consistent with the nearly perpendicular alignment of $H_{\rm int}$ and $V_{zz}$ at the La(2) site\cite{Yashima327}, the powder pattern simulation well reproduces the La(1)-spectra below $T_{\rm SDW}$ with a sufficiently small $H_{\rm int}(\text{La}(1))$ of at most $\sim 10$ mT.
According to the single-crystal NMR spectra\cite{Zhao_NMR}, $H_{\rm int}$(La(1)$^{\rm M}$) $\sim \pm11$~mT was estimated at approximately half of the La(1) sites (La(1)$^{\rm M}$), while it exhibits a negligibly small value ($H_{\rm int} \sim 0$) for the other half (La(1)$^{\rm NM}$).
Among the three stacking patterns, the IS stacking (Ic and I\hspace{-1.2pt}Ic) can be excluded since a sufficiently large $H_{\rm int}$ should emerge at all La(1) sites.
In addition, under the assumption of the dipolar hyperfine field ($H_{\rm hf}^{\rm dip}$) arising from Ni moments of $M_{\rm Ni}$($\parallel a$) $\sim$ 0.5~$\mu_{\rm B}$ ($\approx$ 0.42 - 0.67~$\mu_{\rm B}$ reported previously \cite{Chen_muSR,Yashima327,Plokhikh,Chen-single-neutron}), the $H_{\rm int}$ at all La(1) sites is expected to be approximately 30 mT in the case of the IS stacking. 
These facts are inconsistent with the observation, even though we have neglected the possible transferred hyperfine field $H_{\rm hf}^{\rm tr}$ arising from the two neighboring Ni spins. 
This term would contribute to the same positive direction as the $H_{\rm hf}^{\rm dip}$. 
The FM stacking (Ib and I\hspace{-1.2pt}Ib) is also excluded. 
For the FM stacking (Ib), the $H_{\rm int}$ is indeed almost canceled out at half of the La(1) sites (La(1)$^{\rm NM}$), however, at the remaining half of the sites (La(1)$^{\rm M}$), the $H_{\rm int}$ from four neighboring ferromagnetic Ni spins is estimated to be at least $H_{\rm int}$(La(1)$^{\rm M}$) $\sim \pm55$~mT, considering only $H_{\rm hf}^{\rm dip}$ and ignoring the positive contribution of $H_{\rm hf}^{\rm tr}$. Consequently, the expected $H_{\rm int}$ values at these La(1)$^{\rm M}$ sites for the FM and IS stackings are significantly large compared to the observed value.

Thus, we suggest that the most probable state is the canted AFM stacking (Ia or I\hspace{-1.2pt}Ia).
In this state, the canting of the AFM spins induces an appropriate $H_{\rm int}$ not only at all O(1) sites but also at half of the La(1) sites (La(1)$^{\rm M}$ with $\sim \pm 11$~mT) located along the Ni spin channels. Simultaneously, a nearly canceled field ($H_{\rm int}\sim 0$~mT) is maintained at the other half of the La(1) sites (La(1)$^{\rm NM}$) along the Ni spinless or tiny-spin channels. 
Although the canting angle between a pair of interlayer Ni moments has not been precisely determined, it should be sufficiently small, reflecting the weak $H_{\rm int}$ of  $\sim 11$~mT at the La(1)$^{\rm M}$ site.
Consequently, we remark that the $^{17}$O(1)-NMR and $^{139}$La(1)-NMR\cite{Zhao_NMR} results are simultaneously explained by the canted AFM stacking illustrated in Ia or I\hspace{-1.2pt}Ia of Fig. \ref{Spec20K}(b). 
Spin canting in AFM materials often occurs in crystals featuring an asymmetric local structure via the Dzyaloshinskii-Moriya interaction (DMI): It is defined as ${\bf D}_{ij} \cdot ({\bf S}_i \times {\bf S}_j)$, where ${\bf D}_{ij}$ is the DM vector between two adjacent spins $ {\bf S}_i$ and $ {\bf S}_j$.
Because the interlayer Ni--O(1)--Ni bond is bent within the $bc$-plane in this compound, ${\bf D}_{ij}$ is oriented along the $a$-axis. 
This DMI between two adjacent Ni spins via the O(1) site works when the direction of the Ni spins deviates from the $a$-axis. The previous NQR measurement\cite{Yashima327} reported that the angle $\theta$ between $H_{\rm int}$ and $V_{zz}$ tilting moderately from the $c$-axis toward the $b$-axis is close to, but not exactly, $90^\circ$. 
This indicates that the direction of the Ni spins deviates slightly from the $a$-axis, thereby allowing the DMI to be active.  
It is noteworthy that the canted AFM phase of this compound appears below $T_{\rm A}$, which is well below $T_{\rm SDW}$, indicating that the spin reorientation from AFM appears deep within the SDW ordered phase. 
This suggests that the DMI tilts the spins effectively only when the amplitude of Ni moments has substantially developed well below $T_{\rm SDW}$. 

Next, we address the relationship between the SDW order and the interlayer spin coupling between the NiO$_2$ planes in La327.
Theoretically, the in-plane SDW order in La327 is primarily driven by the Fermi-surface nesting between the $\alpha$ and $\beta$ pockets mostly dominated by the $d_{x^2-y^2}$ orbitals with the nesting vector of (1/4, 1/4) in the tetragonal basis\cite{Luo_theoryPRL,YWang_theoryPRB}, which corresponds to (0, 1/2) in the orthorhombic one\cite{Yashima327}.
As for the interlayer spin configuration, it has been known that the magnetic interaction $J_{\perp}$ through interlayer Ni--O(1)--Ni bonding is significantly larger than the in-plane interaction $J_{\parallel}$ through planar Ni--O(3,4)--Ni bonding\cite{Luo_theoryPRL,Sakakibara_PRL,Chen_RIXS}. 
Regarding the spin interaction between NiO$_2$ layers, one might simply expect an AFM spin configuration due to the strong magnetic interaction $J_{\perp}$. 
However, instead of stabilizing the antiferromagnetic SDW order, this large $J_{\perp}$ forces the adjacent Ni-$d_{3z^2-r^2}$ spins along the $c$-axis to form orbital-selective interlayer spin-singlets, which quenches the majority of these spins, as discussed in Sec. B.
Consequently, any interlayer antiferromagnetic correlation is expected to be attributed to two spin components: (i) a tiny fraction of Ni-$d_{3z^2-r^2}$ spins uninvolved in spin-singlet formation and (ii) $d_{x^2-y^2}$ spins dominating the magnetic moment in the SDW order. Since the $d_{x^2-y^2}$ orbitals extend widely within the $ab$-plane but have a small extension along the $c$-axis, the interlayer correlation between $d_{x^2-y^2}$ spins is unlikely to develop significantly. The resulting effective interlayer coupling $J_{\perp}^{\rm eff}$ driving the AFM stacking is expected to be substantially reduced compared to the original large $J_{\perp}$.
This substantial reduction in $J_{\perp}^{\rm eff}$ provides a natural explanation for the canting behavior due to the DMI effect. Because the spin-canting angle is roughly proportional to $D_{ij}/J_{\perp}$ when this ratio is small, an excessively large $J_{\perp}$, which is predicted by theory, would suppress the canting to a level well below experimental detection. Therefore, the quenching of the Ni-$d_{3z^2-r^2}$ spins via the spin-singlet formation seems to be necessary; without this reduction to a smaller $J_{\perp}^{\rm eff}$, the subtle spin canting might have remained obscured.
Namely, the development of orbital-selective spin-singlet formation leads to the disappearance of spins in the $d_{3z^2-r^2}$ orbital strongly connected by $J_{\perp}$.
In fact, the observed magnetic moment of the Ni site is $0.42 - 0.67$~$\mu_{\rm B}$ below $T_{\rm SDW}$\cite{Chen_muSR,Yashima327,Plokhikh,Chen-single-neutron}, which is much smaller than 1.5~$\mu_{\rm B}$ corresponding to the fully polarized moment assumed from 1.5~electrons/Ni, implying the well-developed spin-singlet state in the $d_{3z^2-r^2}$--O(1)--$d_{3z^2-r^2}$ bonds.

Here, we briefly address the anomalous behavior of $K$ observed below $T_{\rm SDW}$.
As shown in Fig.~\ref{Spec20K}(a), the spectral peaks of the O(2) and O(1) sites exhibit a slight downward frequency shift below $T_{\rm SDW}$, reflecting the decrease in $K$; in particular, $K_{\rm spin}$ at the O(2) site decreases to nearly zero.
Although the detailed mechanism is not yet fully understood, this downward shift could be ascribed to the well-developed interlayer spin-singlet formation, which further progresses toward full development upon cooling below $T_{\rm SDW}$.
While magnetic order and spin-singlet formation are inherently competing phenomena, such an anomalous behavior can be compatible owing to
the orbital-selective roles of the individual Ni orbitals, where the $d_{x^2-y^2}$ spins primarily drive the SDW order and the $d_{3z^2-r^2}$ spins govern the interlayer spin-singlet formation.

We also recall that some reports on the transport experiments have pointed out the charge anomaly at around 110--140 K\cite{Wu,Liu2023,Yanan-Zhang_replication,Taniguchi,Liu2024}. 
This anomaly occurs at a temperature close to $T_{\rm A}$. 
Although the charge anomaly at the ideal La(2)$_a$ site could not be explicitly observed across $T_{\rm A}$ in the previous NQR study\cite{Yashima327}, we cannot exclude the possible occurrence of a tiny charge anomaly below $T_{\rm A}$, which might be associated with the further development of spin-singlet formation upon cooling.
If the $d_{3z^2-r^2}$ electrons lose their itinerancy due to the charge anomaly, it may weaken the in-plane hybridization between the $d_{3z^2-r^2}$ and $d_{x^2-y^2}$ orbitals, and consequently may influence their interlayer spin configuration. 
By applying pressure, this charge anomaly is known to be gradually suppressed\cite{Wu,Yanan-Zhang_replication}, in contrast to the increase in $T_{\rm SDW}$\cite{Khasanov,Zhao_NMR}, and hence the pressure dependence of the anomalous interlayer spin coupling state will provide a crucial indication for elucidating the possible relation between the spin-singlet formation and the anomaly at $T_{\rm A}$. 
These issues may provide important insights to reveal the intrinsic phase diagram of the density wave order and high-$T_c$ superconductivity under high pressure in La$_3$Ni$_2$O$_{7}$.

\section{\leftline{S\lowercase{ummary}}}
The $^{17}$O-NMR measurements on the bilayer nickelate La$_3$Ni$_2$O$_7$ have successfully distinguished the inner apical O(1), outer apical O(2), and planar O(3,4) sites.
Below $T_{\rm SDW}$ (= 150~K), the peak for all planar O(3,4) sites becomes markedly broader due to a finite $H_{\rm int}$ from Ni spins, whereas there remain O(2) sites with no (or a negligibly small) $H_{\rm int}$.
These results are consistent with the commensurate SDW order model featuring the single spin-spinless (or large-tiny spin) stripe. 
They are inconsistent with the other proposed density wave models such as the double spin stripe, double spin-charge stripe, and double charge stripe models.
Regarding the interlayer spin configuration, the $H_{\rm int}$(1) at the O(1) site bridging the NiO$_2$ planes is nearly canceled below $T_{\rm SDW}$ down to $T_{\rm A} \sim 115$~K, indicating that the interlayer spin alignment is predominantly antiparallel between adjacent Ni spins. 
Below $T_{\rm A}$, however, the observation of a finite $H_{\rm int}$(1) demonstrates the collapse of this antiparallel interlayer spin alignment, suggesting that the interlayer antiferromagnetic (antiparallel) spin configuration is not so robust in spite of the presence of a significantly strong $J_{\perp}$.
Among the possible candidates, we propose a scenario involving a transition from the collinear AFM state to the canted AFM state, presumably triggered by the DMI effect arising from the asymmetric bonding between adjacent Ni spins along the $c$-axis.
Above all, we emphasize that the local spin susceptibility $\chi_{\rm spin}^{\rm loc}$ at the O(2) site is extremely small, indicating the well-developed interlayer spin-singlet formation in the Ni-$d_{3z^2-r^2}$ orbital bridging the NiO$_2$ planes.
Instead of stabilizing the antiferromagnetic SDW order, the large $J_{\perp}$ forces the adjacent Ni-$d_{3z^2-r^2}$ spins along the $c$-axis to form interlayer spin-singlets.
This prominent orbital-selective interlayer spin-singlet formation quenches the majority of Ni-$d_{3z^2-r^2}$ spins and leads to the substantial reduction in $J_{\perp}^{\rm eff}$, consequently providing a natural explanation for the canting behavior due to the DMI effect, as well.
By revealing the ambient-pressure coexistence of primarily $d_{x^2-y^2}$-derived SDW order and $d_{3z^2-r^2}$-derived interlayer spin-singlet formation through the $\text{Ni--O(1)--Ni}$ bonding orbitals connecting the NiO$_2$ planes, our results shed new light on the orbital-selective nature of the bilayer nickelate La$_3$Ni$_2$O$_7$, offering key insights into which of the two orbitals plays the dominant role in driving high-$T_c$ superconductivity under high pressure.

\section*{Acknowledgements}

This work was partially supported by the Iketani Science and Technology Foundation, Nippon Sheet Glass Foundation, and JSPS KAKENHI Grant No. JP25K00959, JP26K22290, JP24K01333, JP24K00580, and JP25K00951. 
One of the authors (M.K.) was supported by the Research Fellowship for Young Scientists, Grant No. JP25KJ1758.
Two of the authors (H. S. and Y. T.) were supported by World Premier International Research Center Initiative (WPI), MEXT, Japan.


\end{document}